\documentclass[journal=acsphotonics,manuscript=article]{achemso}

\usepackage[version=3]{mhchem} 
\setkeys{acs}{articletitle = true, doi=true}

\author{Martin Podhorský}
\email{martin.podhorsky@tu-berlin.de}
\author{Maximilian~Klonz}
\author{Lux~Böhmer}
\author{Sebastian~Kulig}
\author{Chirag~C.~Palekar}
\affiliation[TU Berlin]
{Institut für Physik und Astronomie, Technische Universität Berlin, Hardenbergstraße 36, D-10623 Berlin, Germany}
\author{Petr Klenovský}
\affiliation[MU Brno]
{Department of Condensed Matter Physics, Faculty of Science, Masaryk University, Kotlářská 267/2, 61137 Brno, Czech Republic}
\alsoaffiliation[Met Brno]
{Czech Metrology Institute, Okružní 31, 63800 Brno, Czech Republic}
\email{klenovsky@physics.muni.cz}
\author{Sven Rodt}
\author{Stephan Reitzenstein}
\affiliation[TU Berlin]
{Institut für Physik und Astronomie, Technische Universität Berlin, Hardenbergstraße 36, D-10623 Berlin, Germany}
\email{stephan.reitzenstein@physik.tu-berlin.de}

\title[An \textsf{achemso} demo]
  {Buried Stressor Engineering for Position-Controlled InGaAs Quantum Dots with Local Density Variation for Integrated Quantum Photonics}

\abbreviations{}
\keywords{Quantum communication, Photonic quantum technologies, Site-controlled quantum dots, Surface-strain tuning, continuum elasticity, $\mathbf{k}\cdot\mathbf{k}$, configuration interaction, multipole-expansion of exchange}

\begin{document}

\begin{tocentry}

    \includegraphics[width=8.25cm]{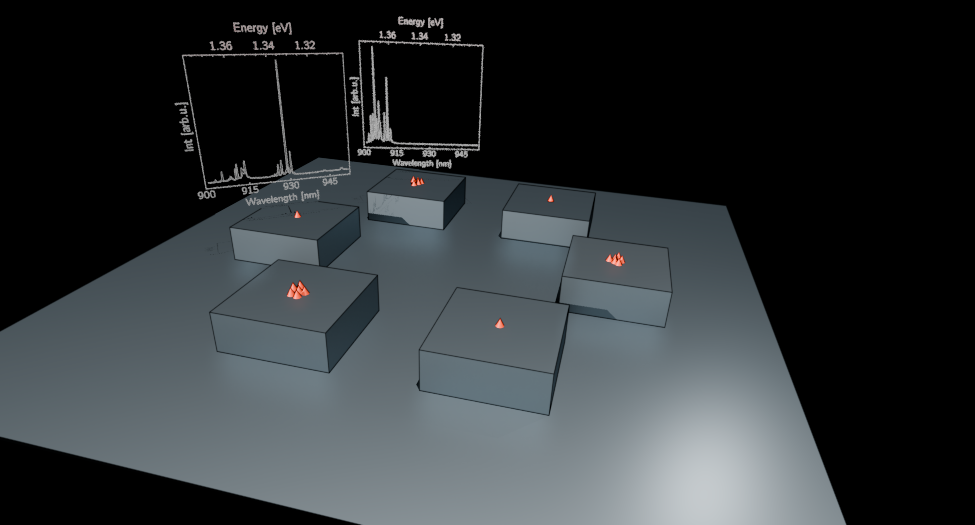}
    \label{For Table of Contents Only}

\end{tocentry}

\begin{abstract}
We report on the monolithic, two-step epitaxial growth of site-controlled InGaAs quantum dots via the buried stressor method with local quantum dot density variation. As a result of high fabrication accuracy, we achieve low lateral displacements of the individual buried stressor apertures of  $
17^{+19}_{-17}$~nm from mesa centers. We provide extensive micro-photoluminescence and cathodoluminescence characterization of the site-controlled quantum dots and give theoretical calculations, explaining the effect of the stressor aperture on the quantum dot emission properties, positioning, and density. We show reproducibility of the nucleation process for apertures of the same size and achieve precisely-positioned, low- and high-density quantum dot nucleation within one active layer growth step. The results presented in this work demonstrate the significant potential of the buried stressor concept in fabricating single photonic chips, simultaneously combining single-photon sources and microlasers featuring different local densities of site-controlled quantum dots, paving the way for highly functional source modules with applications in photonic quantum technology. 
\end{abstract}


\subsection{Keywords}
{\footnotesize Quantum Communication, Photonic Quantum Technologies, Site-Controlled Quantum Dots, Surface Strain Engineering, Continuum Elasticity Theory, $\mathbf{k}\cdot\mathbf{p}$ Method, Configuration Interaction Method}
\normalsize

\section{Introduction}

The field of photonic quantum information technology, including quantum computing and quantum communication, requires on-demand single-photon generation via suitable sources. As such, defect centers in solids, quantum emitters in 2D materials, and semiconductor quantum dots (QDs) have proven their significance as single-photon source (SPS) platforms~\cite{ZhangDefectSPS, DefSchellSPS, 2DPeyskens2019, 2DAzzam, SERAVALLISPS, HeindelSPS}. Among these, self-assembled QDs are of particular interest due to their almost ideal single-photon purity and indistinguishability, combined with a high spontaneous emission rate. These properties enable the highly demanding requirements of advanced quantum optical applications for future photonic quantum technologies to be met, providing a basis for the secure and efficient transmission of quantum information. Furthermore, semiconductor QDs can be employed as an optically active medium in vertically emitting lasers, finding application in areas such as large-capacity optical communication, imaging, biosensing, and neuromorphic computing~\cite{Robertson2022NC, SkalliPNC, HeuserPNC, ChenPNC, XiangPNC, Pan2024VCSEL, ZhouBiosensors, Koyama2014VCSEL}. 

The main drawback of self-assembled QDs is their random nucleation due to the total energy minimization process during surface growth, and the difficulty of achieving control over the nucleation density~\cite{VW}. In order to control QD nucleation, several methods have been developed, including surface patterning techniques, such as tetrahedral recesses or nanohole patterning~\cite{Huang2025Pyramid, PellSCQDPyr, SunnerNanoholes, DirkDroplet, HeynDroplet, SchneiderNanohole, SchneiderNanohole1, BassoDroplet}, or the buried stressor method~\cite{stritt1}. In this work, we present a detailed optical investigation of stressor apertures and positioned QDs, offering a deeper understanding of site-controlled quantum dot (SCQD) growth via the buried stressor method. This includes an analysis of the correlation between stressor aperture width and the emission and nucleation properties of quantum dots. Furthermore, we demonstrate monolithic growth of positioned QDs via the buried stressor method, where we show SCQD nucleation with local control over the QD density via the oxide aperture size on the same chip. This provides a platform for the scalable fabrication of devices, in which low- and high-density QD growth can be realized locally. One example of this is the emitter arrays of quantum light sources and microlasers that are grown within a single epitaxial growth of the active layer. 

\section{Results and Discussion}

\subsection{Sample Growth and Fabrication}

The sample was fabricated by two-step epitaxial growth in a horizontal Aixtron 200/4 MOCVD reactor on a GaAs (001) substrate, using the following group III and group V precursors: trimethylgallium (TMGa), trimethylindium (TMIn), trimethylaluminium (TMAl), arsine (AsH$_3$) and tert-butylarsine (TBA) with hydrogen (H$_2$) as the carrier gas. The growth rates and compositions of the binary and ternary compounds were calibrated from high-resolution X-ray diffraction (HRXD) scans of InGaAs/GaAs, AlGaAs/GaAs, AlAs/GaAs superlattices with Malvern X`Pert$^{3}$ Material Research Diffractometer and from the in-situ reflectance anisotropy spectroscopy measurements via LayTec EpiRAS system.

In the first growth step, a 200~nm GaAs buffer layer is grown after an initial oxide desorption process in an AsH$_{3}$-rich atmosphere, followed by the deposition of 33.5 distributed Bragg reflector (DBR) mirror pairs, consisting of 78.5~nm Al$_{0.9}$Ga$_{0.1}$As and 66.8~nm GaAs layers. Subsequently, a stressor layer stack is grown, comprising a 10~nm Al$_{0.1\rightarrow0.9}$Ga$_{0.9\rightarrow0.1}$As graded layer, a 40~nm Al$_{0.9}$Ga$_{0.1}$As layer, a 30~nm AlAs layer, a 40~nm Al$_{0.9}$Ga$_{0.1}$As layer, and a 10~nm Al$_{0.9\rightarrow0.1}$Ga$_{0.1\rightarrow0.9}$As graded layer. Finally, a 80 nm GaAs  layer is deposited. The temperature at which all layers are grown in this step is 700 °C. 

\begin{figure}[h!]
    \centering
    \includegraphics[width=1.\linewidth]{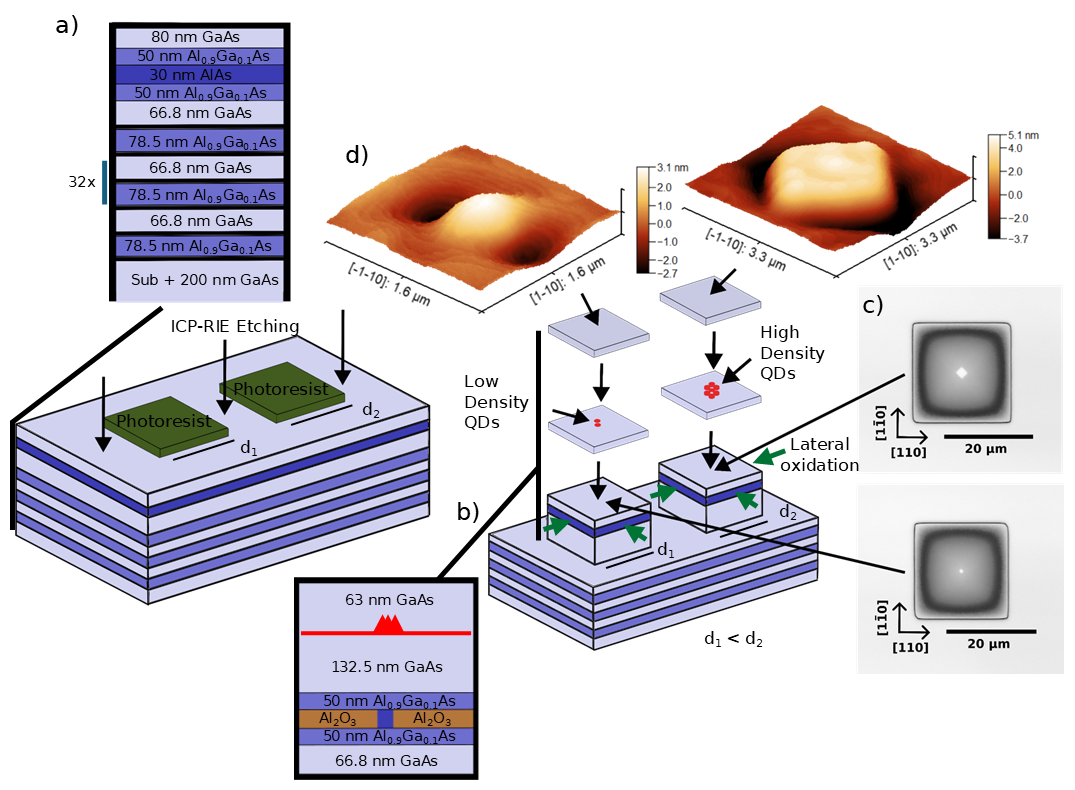}
    \caption{Schematics of the sample fabrication. a) Layer design and illustration of the ICP-RIE step to uncover the AlAs layer. b) Illustration of the in-situ lateral oxidation of the AlAs layer which results in change of the surface strain profile. c) CLSM images showing oxide apertures, where the different aperture sizes can be observed for smaller (bottom) and larger (top) mesa. d) Self-aligned defect which can be resolved on the surface via AFM.}
    \label{schema}
\end{figure}

The sample is then removed from the reactor and processed in a clean room. First, the sample is spin-coated with 0.5~$\mu$m grade AZ nLOF 2070 photoresist at 4000 revolutions per minute. Subsequently, the sample is baked at 110 °C for 60~s. This is followed by the electron beam lithography (EBL) patterning of square mesas with various sizes in the range of 20.3--24.9~$\mu$m with a step size of 0.1~$\mu$m. After exposure, the sample is baked again at 110 °C for 60~s and developed in AZ 726 MIF developer for 15~s. After development, the sample is etched anisotropically using a Sentech~SI~500 inductively coupled plasma reactive ion etching (ICP-RIE) system to uncover the buried AlAs layer, after which it is cleaned in a 70 °C dimethyl sulfoxide (DMSO) solution. Further details of the etching process can be found in the Supporting Information (SI), section S6. The initial growth and fabrication steps are schematically represented in Fig.~\ref{schema}~a). After any residuals are removed, the sample is laterally oxidized in an in-situ oven with water vapor as the oxidizing agent and nitrogen (N$_{2}$) as the carrier gas (Fig.~\ref{schema}~b)). This step is similar to the lateral oxidation step in vertical-cavity surface-emitting laser (VCSEL) processing. However, here the oxidized layer serves not to confine current in the active region, but to modify the surface strain profile at the QD growth surface in order to induce site-controlled growth. A cross-section of resulting strain distribution can be found in K. Gaur, et al.\cite{BSrev}. The oxidization is performed at 405 °C as, for the temperatures closer to 390 °C, lower oxidation speed anisotropy between the $\left<100\right>$ and $\left<110\right>$ directions is present for similar oxidation conditions~\cite{VaccaroOxi}. This ensures higher symmetry of the oxide apertures, especially for smaller aperture sizes. The process is stopped once no apertures can be observed via an optical microscope in the smallest patterned mesa (20~$\mu$m). The aperture sizes are then defined by the step size (0.1~$\mu$m) between the mesa sizes. The total oxidation time was 15 minutes. The remaining larger mesas then accommodate oxide apertures of different sizes, which determine the behavior of the SCQD nucleation. In Fig.~\ref{schema}~c), we show examples of a mesa with smaller (bottom) and larger (top) size, measured with confocal laser scanning microscope (CLSM). The width of the oxide apertures can be obtained from these images via fitting with a step function, as explained in detail in the SI Section S2. When discussing the oxide aperture dimensions, we refer to the widths obtained via this method. The difference in aperture size between the smaller~(d$_{1}$) and larger~(d$_{2}$) mesas can be seen. The variety of mesa sizes, and therefore the oxide aperture sizes, creates different surface strain profiles above the stressor apertures~\cite{ImadBS}. The oxidized sample is then cleaned in de-ionized water prior to the second growth step. 
In the second growth step, a 55.2~nm-thick GaAs layer is grown at  700 °C, after oxide desorption. The reactor temperature is then decreased to 500 °C, followed by the growth of 2.7~monolayers~(ML) of In$_{0.52}$Ga$_{0.48}$As with the growth rate of 0.2~ML/s. The QD layer thickness is above the critical thickness for Stranski-Krastanov nucleation, and the V/III ratio during the active layer growth was~1.5. The combination of the deposit's low thickness, lower growth rate and lower V/III ratio results in very low planar QD densities~\cite{Hsieh_2006LDQD, Richter_2010LDQD, PETROFF}. As a consequence, very high selectivity for the QD nucleation above the stressor aperture can be achieved. The deposition is followed by a 50~s growth interruption time with no flux of the V and III precursors. The QDs are then overgrown with nominally 1~nm thick GaAs capping layer. After capping, In desorption step is performed by increasing the reactor temperature to 615 °C for 240~s. This step is introduced with the intent to achieve higher QD uniformity. Once the desorption step is completed, the final 63~nm GaAs layer is grown at that temperature. In Fig. 1 d), atomic force microscopy (AFM) surface profiles after the growth of the final GaAs layer are shown for smaller (left) and larger (right) mesas. We observe the formation of a self-aligned defect above the QDs, as reported by C.-W. Shih et al. [58]. Different morphology of the defect can be resolved for both smaller (left) and larger (right) mesa. This distinction is a result of the individual underlying surface strain profile.

\subsection{Experimental and Theoretical Results}

In order to perform a systematic, quantitative study of the effect of oxide apertures on QD nucleation, it is necessary to determine the relationship between the mesa size and oxide aperture size, as well as the distribution of oxide aperture sizes on the wafer. For this purpose, a reference sample was grown and processed under conditions that were identical in all respects, with the exception of a reduced number of bottom DBR mirror pairs. As shown in work of K. Gaur et al. [27], the surface strain is influenced by the thickness of the AlAs stressor layer and the thickness of the deposited layers between the QD surface and the stressor. The layers below the stressor are not expected to contribute to the surface strain profile. Similarly, since the aperture sizes are given by the lateral oxidation of the AlAs layer, no DBR dependent change in the oxidation properties are to be expected if the fabrication conditions are kept the same.

\begin{figure}[h!]
    \centering
    \includegraphics[width=1.0\linewidth]{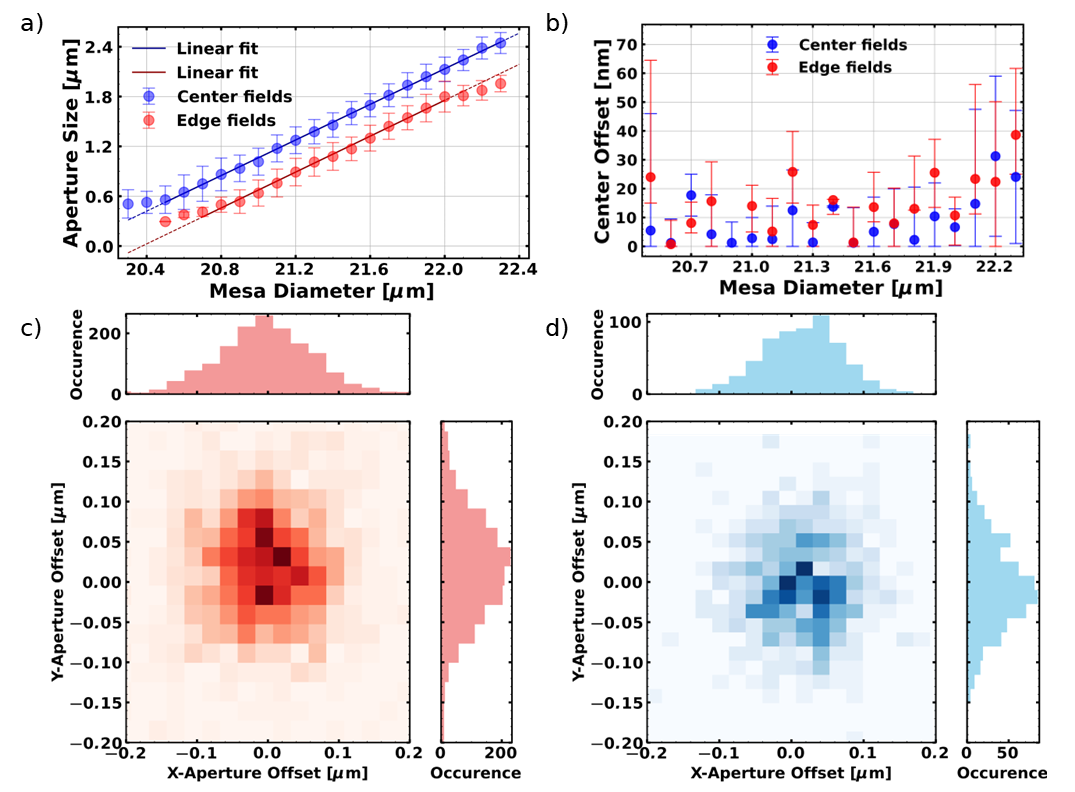}
    \caption{CLSM investigation of stressor apertures. a) Mean aperture sizes for individual mesa sizes are fitted with linear functions. b) Absolute aperture offset from the mesa center showing very low aperture displacement. Panels c) and d) show the histograms aperture offsets on sample center and edge, respectively. }
    \label{CLSM_comp}
\end{figure}

In Fig.~\ref{CLSM_comp}~a), the evolution of the oxide aperture size with the mesa size is shown. To demonstrate the differences caused by fabrication, we present a comparison of aperture evolution at two separate representative positions on the wafer, i.e. the center (midpoint of the sample) and the edge (near the sample boundary), respectively. The sample studied was a 1/4 piece of a 2 inch wafer, the distance between the measured fields was approximately 1~cm. The two datasets represent the mean aperture sizes of 4$\times$4 patterned fields, corresponding to a selected area of roughly $110$~mm$^{2}$ on the wafer. The aperture sizes were investigated using CLSM. First, images of mesas with various sizes were acquired and the aperture widths were determined by fitting the aperture region with a step function. Furthermore, the center of the aperture is determined as the central point of the fitted step function, offset from the actual mesa center. The center of the mesa is found relative to the mesa border through edge detection in the CLSM image. These procedures are explained in further detail in the SI, section S2. 
The disparity in aperture size between the center and edge positions of the individual mesas in Fig.~\ref{CLSM_comp}~a) is due to inhomogeneous oxidation in the in-situ oven. This results in larger oxide apertures in smaller mesas at the sample edge compared to the center. Other aspects, such as thickness inhomogeneity from template growth and photoresist thickness inhomogeneity from spin-coating, also influence the outcome of the fabrication process. In both cases, there is a clear linear dependence for aperture sizes of 500~nm, with slopes of 1.08(2) and 1.07(1) for the center and edge pieces, respectively. These values are consistent with each other within the uncertainty, reflecting the robustness of the oxidation process. It is noteworthy that, despite the variation in aperture size across the wafer, the positioning of the apertures relative to the mesa center remains unchanged, as can be seen in the histograms of the lateral and vertical offsets of the measured apertures at the edge and center of the sample. The respective histograms are shown in Fig.~\ref{CLSM_comp}~c) and Fig.~\ref{CLSM_comp}~d). The points and error bars in Fig.~\ref{CLSM_comp}~b) represent the mean value of the measured aperture offset from the center of mesa for a given mesa size, along with its standard deviation, respectively. As illustrated, the absolute lateral displacement of the oxide aperture remains below 60~nm for the individual mesas with an average value of $
17^{+19}_{-17}$~nm. 

To better understand the nucleation behavior and emission properties of SCQDs, we conducted extensive cathodoluminescence (CL) and micro-photoluminescence ($\mu$--PL) investigations. First, we focused on the statistical evaluation of the emission properties obtained from $\mu$--PL measurements. Similar to the aperture investigation, we have selected neighboring fields on two positions on the wafer.  

\begin{figure}[h!]
    \centering
    \includegraphics[width=1.0\linewidth]{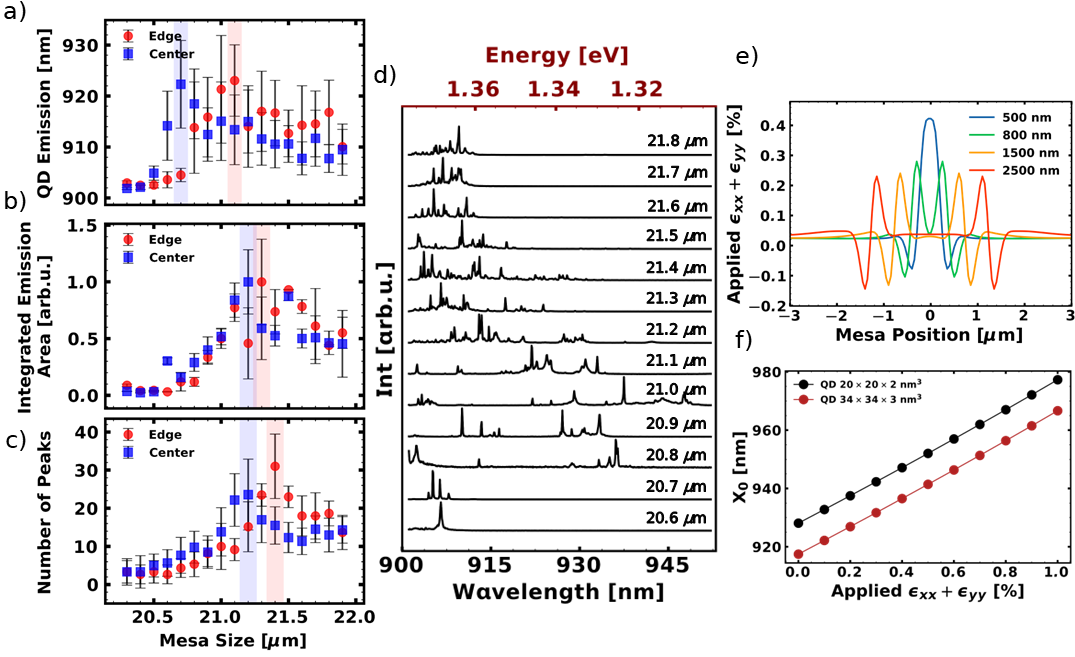}
    \caption{$\mu$--PL investigation of the QD emission and theoretical calculations of the biaxial strain. Panels a), b), and c) depict the mean QD emission wavelength, the mean integrated QD emission area, and the mean number of QD peaks, respectively. The highlighted regions show the maximum values. d) Selected spectra for various mesa sizes from a single patterned edge field. e) Continuum elasticity theory calculation of the biaxial surface strain profile for different aperture sizes. f) Calculated exciton wavelength shift with respect to the applied biaxial. All QD spectra were excited with a tunable pulsed laser at 890~nm. The excitation power was kept constant at 1.5~$\mu$W.}
    \label{uPL}
\end{figure}
From the measured spectra, we can obtain the mean QD emission. This is defined as the mean wavelength of all the peaks identified in the spectra taken for mesas of a given size. These spectra are also used to calculate the mean number of QD peaks. The integrated emission is obtained by integrating the area under the emission spectrum within the range of 905-–960~nm using Python's built-in integrating functions. The peak detection procedure, emission area integration and evaluation are discussed in detail in the SI, section S4. Python's built-in \texttt{find\_peaks()} function is used to detect the lines in each spectrum, which are then fitted with a Lorentzian. Thus, parameters for each measured mesa size are extracted, and a clear trend emerges in both the wavelengths of QD emission, and the number of emission lines observed. We assume mixtures of neutral excitons (X), positive (X$^{+}$) and negatively charged excitons (X$^{-}$). Biexcitons and excitonic complexes with higher number of carriers can be neglected because of the low excitation power, well below saturation of X, X$^{+}$, and X$^{-}$. The statistical evaluation in Fig.~\ref{uPL}~a) incorporates lines of all the mentioned excitonic complexes. We assume a correlation between the number of QDs and the appearance of X, X$^{-}$, and X$^{+}$ lines. The details can be found in Section S4 of the SI.
The dependence of the extracted quantities on the mesa sizes is shown in Figs.~\ref{uPL}~a--c), respectively. In Fig.~\ref{uPL}~a), we observe a discrepancy between the largest shift of the emission wavelength between the center (20.7~$\mu$m) and edge (21.1~$\mu$m) positions. We contribute this to the inhomogeneity of the oxide apertures induced by the sample fabrication, namely by the oxidation process. This results in the aperture size in the center of the sample being larger than on the edge for equivalent mesa sizes. As such, smaller apertures on the center (edge) will be present in smaller (larger) mesas. A qualitative comparison can be made with the CLSM measurements of the aperture sizes, shown in Fig.~\ref{CLSM_comp}~a). This also gives rise to the difference between the maxima of the integrated emission area (Fig.~\ref{uPL}~b)) and the maximum of QD peaks (Fig.~3~c)), which can be observed for mesa sizes around 21.1~$\mu$m and 21.4~$\mu$m for sample center and sample edge, respectively. Here, these extrema are used to indicate the highest SCQD density. In SI, section S4, we present the mean QD emission wavelength, the mean integrated QD emission area, and the mean number of QD peaks shown in Fig.~\ref{uPL}~a)--c) also a function of aperture sizes, based on the CLSM measurements. 

Since the surface strain profile exhibits a significant unimodal maximum for smaller aperture sizes of around 500~nm (see Fig.~\ref{uPL}~e)), the SCQDs experience the largest applied strain. As can be seen from the calculated wavelengths in Fig.~\ref{uPL}~f), the emission shift between the unstrained dots ($\epsilon_{xx} + \epsilon_{yy}$ = 0.0~\%) and the dots experiencing strain from a 500 nm aperture ($\epsilon_{xx} + \epsilon_{yy}$ = 0.4~\%) corresponds to roughly 20~nm. These results are obtained using continuum elasticity and eight-band ${\bf k}\cdot{\bf p}$ calculations with Coulomb interaction multi-particle correction obtained using the configuration interaction (CI) method, shown in Figs.~\ref{uPL}~e)~and~f). Details on the numerical calculations can be found in the SI, section S1. The QD dimensions in the simulations correspond both in terms of morphology and composition to typical planar Stranski-Krastanov InGaAs QDs \cite{HONG_QD_morphRev}. The maximum mean QD wavelengths displayed in Fig.~\ref{uPL}~a) demonstrate a comparable wavelength shift from the lines exhibiting the shortest emission wavelengths. Consequently, for these mesa sizes, the aperture size at the edge and center positions of the sample must be approximately 500 nm. 

In Fig.~\ref{uPL}~e), we see that with increasing aperture width, the strain profile undergoes a change from unimodal to bimodal distribution. At the same time, the magnitude of tensile strain decreases from 0.4~\% to 0.2~\% as the aperture width increases from 500~nm to 2500~nm. This decrease in applied strain with increasing aperture width results in a reduction in the shift in the wavelength of QD emission, as can be seen in Fig.~\ref{uPL}~f). On the other hand, the increase in aperture width provides a substantially larger area for the QDs to nucleate. We find that the optimum for the mesa size difference between low and high QD densities is approximately 0.4--0.5~$\mu$m. This value is based on the difference between the maximum of the mean QD emission wavelength (Fig.~\ref{uPL}~a), indicating smaller aperture and lower QD densities, and the maxima of the integrated emission area and the number of peaks, exhibiting the highest QD density (Fig. \ref{uPL}~b) and c)). In Fig.~\ref{uPL}~d), we show selected spectra for several mesa sizes from a single patterned field on the edge piece of the wafer. The measurements show a noticeable wavelength shift of the QD lines in the low-density spectra for a mesa size of 21.0~$\mu$m (corresponding to the aperture size of 500 nm), caused by the applied biaxial strain. The number of emission lines then increases, reaching a maximum at a mesa size of 21.4~$\mu$m (corresponding to aperture size of 890 nm). This is accompanied by a gradual decrease in the QD emission wavelength and number. 

\begin{figure}[h!]
    \centering
    \includegraphics[width=1.0\linewidth]{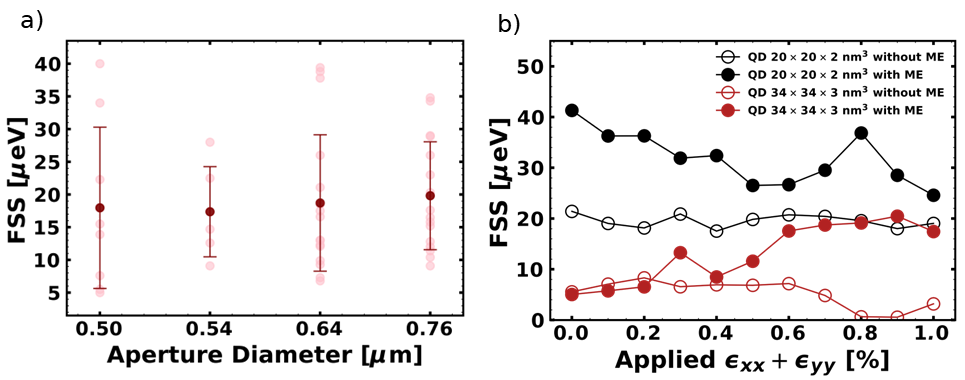}
    \caption{A comparison of measured and simulated FSS. a) FSS dependency on the aperture size, measured from low-density spectra of the SCQDs. b) Calculated FSS values for QDs with dimensions of $20\times20\times2$~nm$^{3}$ and $34\times34\times3$~nm$^{3}$. The abbreviation ME in b) refers to the use of the multipole expansion of the exchange interaction in our calculations~\cite{Takagahara2000,Krapek2015}. The slight cusps and dips in the dependencies in b) are due to numerical errors in the calculations.}
    \label{FSS_meas}
\end{figure}

Additionally, we performed polarization-dependent measurements to explore whether the size of the mesa, and thus the surface strain induced by the oxide aperture, affects the fine structure splitting (FSS) of the nucleated SCQDs. An example of a measurement can be found in the Supplementary Information (SI), section S4. It is anticipated that the FSS will change in response to any distortion in the symmetry of the dot induced by higher values of anisotropic in-plane strain from the stressor aperture~\cite{Diana_PK}. For an isotropic in-plane strain distribution, as assumed in our theoretical calculations, the FSS values are expected to remain unchanged~\cite{Wang_FSS}. Figs.~\ref{FSS_meas}~a) and b) illustrate the measured FSS values for different aperture sizes and the theoretical excitonic calculations of FSS dependency on the applied biaxial strain using CI method. 

\begin{figure}[h!]
    \centering
    \includegraphics[width=1.0\linewidth]{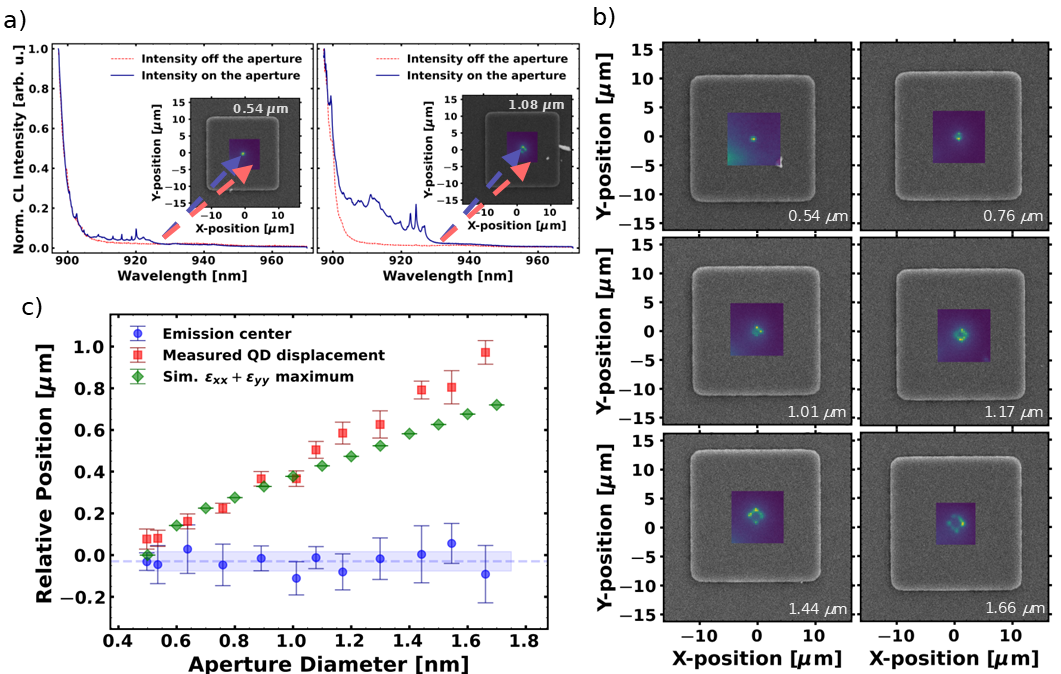}
    \caption{SCQD investigation via CL mapping. a) Selected CL intensity profiles taken on and off the aperture. The insets show the corresponding SEM images overlaid with respective CL maps, for aperture sizes of 0.54~$\mu$m (20.8~$\mu$m mesa) and 1.08~$\mu$m (21.3~$\mu$m mesa). Panel b) SEM images overlaid with their respective CL maps for increasing aperture sizes. c) Dependency of the CL-measured QD displacement (red), CL-measured emission center (blue), and the simulated positions of the tensile strain maxima (green) for increasing aperture sizes. }
    \label{CL}
\end{figure}

The mean value of the measured FSS of the low-density lines for mean aperture sizes of 0.50--0.76~$\mu$m (corresponding to mesa sizes of 20.7--21.0~$\mu$m) is 19.5(9.6)~$\mu$eV. The aperture sizes are based on the CLSM measurements. This value is comparable to those observed in self-assembled InAs/GaAs QD samples grown on the same substrate orientation~\cite{TreuFSS, StevensonFSS, HoegeleFSS}. However, it is significantly larger than the homogeneous emission linewidth in the range of 1~$\mu$eV, meaning that temporal post-selection or post-growth treatment is needed to generate polarization-entangled photon pairs from the biexciton-exciton cascade~\cite{HuwerTF, KupkoQKDImp,HuberREVIEW, VajnerREVIEW}. As can be seen in Fig.~\ref{FSS_meas}~a), there is almost no change in the FSS splitting within this range. Therefore, we conclude that the applied biaxial strain has little to no effect on the QD or substrate anisotropy. This conclusion is reinforced by theoretical excitonic calculations using the CI method, which show similar FSS values to those observed in the experiment (see Fig.~\ref{FSS_meas}~b). Here, we show FSS calculations for two typical truncated-cone-shaped InGaAs QD sizes,~i.e. one with basis size of 20~nm and height of 2~nm and the other with 34~nm basis and 3~nm height. Moreover, for each aforementioned QD structures in Fig.~\ref{FSS_meas}~b), FSS for the case when the multipole expansion of the exchange interaction~\cite{Takagahara2000,Krapek2015} was (was not) considered in CI by full (open) symbols. The principle of multipole interaction is discussed in detail in works of T. Takagahara et al. [36] and V. Křápek et al. [37]. The method provides correction to Coulomb interaction calculations in InGaAs QDs by expansion of Kane’s parameter. We can deduce from Fig.~\ref{FSS_meas}~b) that (i) the smaller dot (20~nm base and 2~nm height) shows larger FSS magnitudes than the larger dot (34~nm base and 3~nm height) and (ii) incorporation of multipole expansion of exchange increases FSS even further for each dot size~\cite{Seguin2005}. Noticeably, while for the case without multipole expansion FSS does not change appreciably with increase of tensile biaxial strain, when that is switched on a slight overall change of FSS with applied strain is observed. Interestingly, in the latter case, FSS increases (decreases) with applied biaxial strain in the case for the larger (smaller) dot. Nevertheless, all aforementioned computed magnitudes of FSS lie within the error bars of the experimentally observed values in Fig.~\ref{FSS_meas}~a). Our results indicate an isotropic in-plane strain distribution of the smaller oxide apertures below $0.8$~$\mu$m by showing no change of the FSS with respect to increasing aperture size. Considering QD nucleation, where the QDs are highly similar with respect to their morphology and composition. For larger aperture sizes, with square-like arrangement of QDs (as can be seen later in Fig.~\ref{CL}~c)), the aperture and the strain becomes in-plane anisotropic, with fourfold symmetry (C$_{4}$), as is shown in the Supporting Information of Ref. [29]. Here, it is impossible to perform such an FSS analysis due to the presence of increased number of QD lines. In order to measure the strain distribution of the apertures in detail, high-resolution techniques, such as X-ray nanodiffraction, need to be implemented~\cite{HraudaXRDnano, MMXRDnano}.   

Fig.~\ref{CL}~a), shows exemplary low- and high-density CL intensity spectra for a selected mesa with smaller (0.54~$\mu$m) and larger (1.08~$\mu$m) aperture size. These aperture sizes are based on CLSM measurements. The insets depict the corresponding SEM images overlaid with the acquired CL maps. The maps and spectra show that the emission is centered in the middle of the mesa. As we increase the aperture size, we observe a change in the distribution of the nucleated QDs, as can be seen in Fig.~\ref{CL} b). This can be attributed to the transition from unimodal to bimodal strain distribution. As the aperture becomes wider, the QDs are positioned at its edges, specifically at the tensile strain maximum positions. This preferential nucleation of dots at the tensile strain maximum positions is caused by the reduced lattice mismatch between the deposit and the substrate at these positions. This results in a local decrease in free energy for QD growth, favoring QD nucleation at positions of tensile strain maximum. Consequently, a transition from a point-like arrangement of QDs in smaller apertures to a square-like arrangement in larger ones can be observed, with the arrangement becoming increasingly uneven as the aperture size increases.

The QD displacement and emission centers are calculated using fits to radial cuts of the intensity distribution. The mesa center is defined as the center of the square formed by the intersections of lines parallel to its edges. Radial cuts of length 5 µm are extracted at 10° intervals across the mesa center to obtain the intensity profiles, which are then individually fitted with a Gaussian or double Gaussian function based on the shape of the acquired intensity profile. The positions of the maxima obtained from the fit provide the QD displacement from the mesa center as a Euclidean distance. We also consider the maximum of the unimodal distribution to be the emission center. For the bimodal distribution, the emission center is defined as the center of mass between the two maxima. The fitting procedure is described in detail in the SI, section S3.

In Fig.~\ref{CL}~c), we show the QD displacement and QD emission offset from the mesa center of one patterned field as a function of aperture sizes, depicted as red and blue data-points, respectively. The simulated positions of the tensile strain maxima are shown in green. The QD emission center exhibits a nearly constant value of -30(47)~nm. Namely in the case of low-density spectra, such QD displacement has little to no impact on the photon extraction efficiency (PEE) of the QDs~\cite{Descamps_valZwiller}. The QD displacement shows dependence on the simulated position of the tensile strain maxima, with strong agreement between the experimental and theoretical results for smaller apertures of up to 1.17~$\mu$m. For larger aperture sizes, we notice a discrepancy between the simulated tensile strain values and the measured QD positioning. This can be attributed to lowered control over the QD positioning due to the lowered tensile strain value and to the anisotropy of the oxide apertures (shown in Ref. [29]).

In practical applications, QDs are usually coupled to a fabricated resonant cavity, such as a micropillar cavity, a circular Bragg grating (CBG) cavity or a photonic crystal (PhC) cavity~\cite{LiuMicropillar, Moczała-DusanowskaCBG, OhtaPhCNanocavity}. Alternatively, they can be integrated into a waveguide~\cite{UppuSDWaveguide}. In general, any QD displacement can negatively impact the position-dependent light-matter coupling strength of cavity-coupled emitter systems, affecting the Purcell effect and the PEE. For example, displacements of more than 50 nm have been shown to have a highly detrimental effect on the PEE of CBG, nanobeam or photonic crystal waveguide cavities~\cite{MadigawaDonges}. Furthermore, in the case of CBG cavities, the Purcell factor, is also considerably decreased~\cite{RickertACS}. On the other hand, the emission properties of QD-micropillars with a larger lateral extent of the confined cavity modes in the range of several 100 nm are less prone to emitter displacements~\cite{SR_MicropillarRev}.  

Density and position control can be exploited to take advantage of the properties of both high- and low-density QDs simultaneously. This can all be achieved within a single active layer overgrowth.
To this end, we created hexagonal arrays involving mesas of two different sizes arranged in an alternating pattern. The size difference between the smaller and larger mesas within each hexagon was set to 0.5~$\mu$m. In such a configuration, it is possible to create a device with three reproducible low- and high-density QD nucleation areas for the future realization of single-photon sources and microlasers, respectively. In this approach, each pair of quantum and classical channels comprises a low-density mesa for quantum key generation and a high-density mesa to provide the communication channel. This way one would be able to implement three secure data transmission channels in parallel, effectively increasing the achievable data rate by a factor of three when using six-core fibers in a hexagonal configuration, while avoiding crosstalk between the quantum and classical signals.~\cite{Dynes:16}. 

\begin{figure}[h!]
    \centering
    \includegraphics[width=1.0\linewidth]{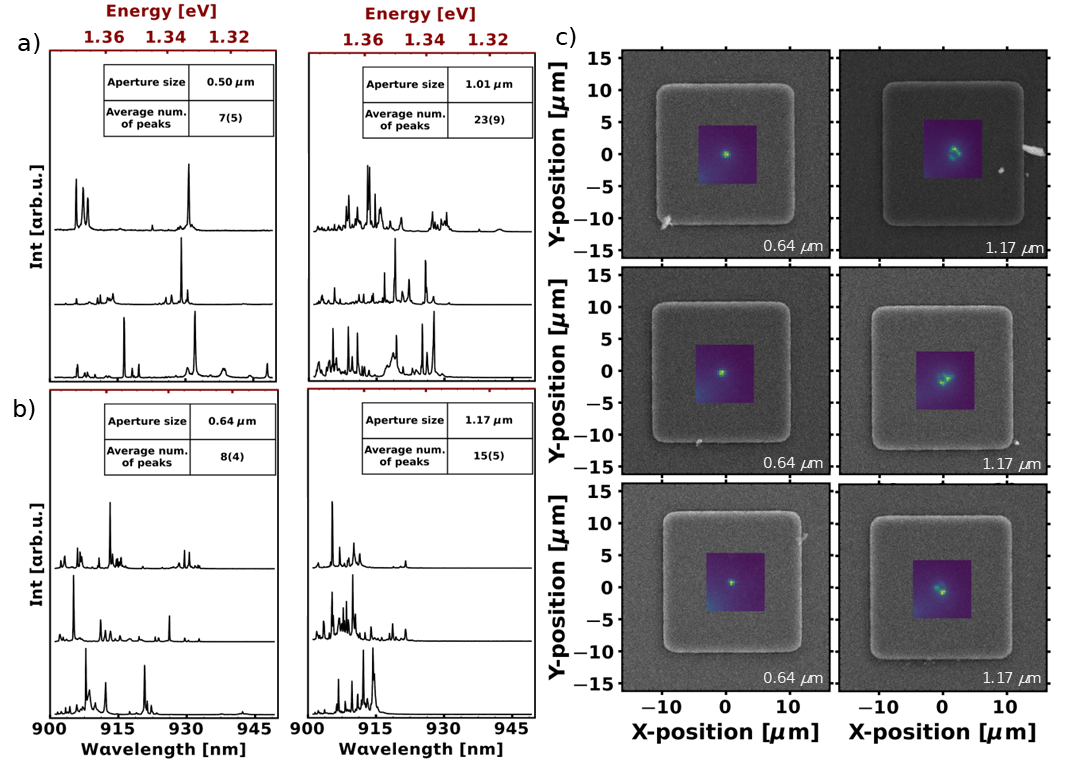}
    \caption{Investigation of hexagonal arrays via $\mu$-PL spectroscopy and CL mapping. a) QD spectra from a single hexagon of with 0.50~$\mu$m and 1.04~$\mu$m apertures (20.7~$\mu$m and 21.2~$\mu$m mesa, respectively). b) QD spectra from a single hexagon of with 0.71~$\mu$m and 1.35~$\mu$m apertures (20.9~$\mu$m and 21.4~$\mu$m mesa, respectively). c) SEM images overlaid with measured CL maps for a hexagonal array, the aperture sizes are identical to b)}
    \label{uPL1}
\end{figure}

In order to highlight the uniformity of SCQD nucleation properties within hexagonal arrays, we present measured hexagons for different aperture sizes in Fig.~\ref{uPL1}. Here, we show spectra taken with $\mu$--PL spectroscopy from hexagonal arrays with aperture sizes of 0.50~$\mu$m and 1.01~$\mu$m (mesa sizes of 20.7~$\mu$m and 21.2~$\mu$m, respectively) for smaller and larger mesa in Fig~\ref{uPL1}~a), and with aperture sizes of 0.64~$\mu$m and 1.17~$\mu$m for smaller and larger mesas (mesa sizes of 20.9~$\mu$m and 21.4~$\mu$m, respectively) in Fig.~\ref{uPL1}~b). For the case of hexagonal array with 0.64~$\mu$m and 1.17~$\mu$m aperture sizes, we show SEM images overlaid with CL maps in Fig.~\ref{uPL1}~c).  We observe similar emission characteristics for apertures of the same size. For apertures of sufficiently small sizes (aperture sizes of 0.50~$\mu$m and 0.71~$\mu$m), isolated emission lines appear in the $\mu$--PL spectra. In section S5 of the SI, to prove single-photon emission, we present second-order autocorrelation measurements of low-density spectra for a selected hexagon, alongside an example of a lifetime measurement.

In the CL maps, the emission is always point-like and centered in the middle of the mesa. Here, the tensile strain maximum is unimodal, which allows for the precise nucleation of low-density QDs. For larger aperture sizes (aperture sizes of 1.01~$\mu$m and 1.17~$\mu$m), as the strain profile transitions from unimodal to bimodal, the QD spectra exhibit an increased number of lines. The stressor now provides a larger total strained area for QD nucleation. In this instance, the QDs are arranged in a square-like configuration.  Consequently, we demonstrate the successful modulation of QD density via surface strain modulation with the buried stressor method.

Quantum light emission and laser emission of SCQDs grown via buried stressor method have been reported on in previous, separate works \cite{GrosseAPL, SCQD_las}. Future work will focus on introducing a suitable resonant cavity design to combine these properties in one sample. If successful, low-output, high-$\beta$ lasing could be achieved with a limited number of QDs, which has already been reported on with use of non-positioned QDs by C. Gies et al.~\cite{Gies_SR} This would extend the possible applications of SCQDs as optical gain medium for microlasers into various fields, from coherent excitation sources in the field of quantum nanophotonics~\cite{SR_LTLaser, KreinbergLSA, MunnellyACS}, to biosensing, chemical detection, nonlinear optical microscopy and low-power on-chip optical data communication~\cite{Shambat2013, Loncar, Nakayama2007, Hill2014}. 
\section{Conclusions and Outlook}

We have shown a scalable method of site-controlled nucleation of InGaAs QDs via the buried stressor method, with high control over the local nucleation density in emitter arrays. 
First, we characterized the oxidation process via CLSM. Despite the robust fabrication method, we show high precision of the stressor aperture placement. We achieve a very low lateral displacement of the individual buried stressor apertures of  $
17^{+19}_{-17}$~nm. We establish the dependency of the aperture size on the mesa size. 
Subsequently, we perform statistical $\mu$--PL and CL characterization on the sample, supported by theoretical calculations, explaining the effect of the stressor aperture on the QD emission properties and on the QD nucleation. We show strong agreement between the experimental and theoretical results.
Lastly, we report on the reproducibility of the nucleation properties. We show $\mu$--PL and CL measurements collected for given aperture sizes. These measurements reveal that the QD and CL emission spectra are similar for the same aperture size.
The results presented in this work provide a deeper understanding of the growth of SCQDs via the buried stressor method, particularly with regard to the correlation between stressor aperture width and the emission and nucleation properties of the QDs. This work will lay the groundwork for future applications in the field of quantum photonics, where monolithic devices that combine classical vertical-cavity surface-emitting lasers (VCSELs) and quantum light emitters are integrated onto a single chip. This was achieved by exploiting the controlled growth of QDs with well-defined high and low densities. When paired with a suitable cavity system, such platforms provide a basis for fiber-coupled devices that can enhance fiber-based quantum networks, enabling quantum key distribution (QKD) and quantum repeaters based on entanglement distribution.

\begin{acknowledgement}

The authors thank Kathrin Schatke, Praphat Sonka, Heike Oppermann, and Stefan Bock for the technical assistance. The authors further thank the project partners from VI Systems GmbH, Fraunhofer IZM, and JCM GmbH for their support.   

\section{Funding Sources}

The authors acknowledge financial support from the German Federal Ministry of Education and Research (BMBF) via the project MultiCoreSPS (Grant No. 16KIS1819K), from the German Research Foundation (DFG) via grant INST 131/795-1 FUGG, together with the funding from the European Innovation Council Pathfinder program under grant agreement No 101185617 (QCEED) and the funding by the Institutional Subsidy for the Long-Term Conceptual Development of a Research Organization granted to the Czech Metrology Institute by the Ministry of Industry and Trade of the Czech Republic and by the project Quantum Materials for applications in sustainable technologies, CZ.02.01.01/00/22-008/0004572.

\end{acknowledgement}

\begin{suppinfo}

Principle of the continuum elasticity theory; methodology of the oxide aperture determination with measurement details; figure of the oxide aperture measurement procedure; methodology of the CL measurements with the emission offset determination procedure; figure of the CL emission offset determination procedure; figure of the QD emission offset; methodology of the statistical u-PL measurements with the measurement details; figure of the the mean QD emission wavelength, the mean integrated QD emission area, and the mean number of QD peaks with respect to the mesa and aperture size; details on the time-resolved measurements; figure of power- and polarization-series of a selected SCQD spectrum with the corresponding fits, together with lifetime measurements of X and XX lines; second-order autocorrelation measurements of 3 low-density mesas from a selected hexagonal array; details on the ICP-RIE etching process

\end{suppinfo}


\providecommand{\latin}[1]{#1}
\makeatletter
\providecommand{\doi}
  {\begingroup\let\do\@makeother\dospecials
  \catcode`\{=1 \catcode`\}=2 \doi@aux}
\providecommand{\doi@aux}[1]{\endgroup\texttt{#1}}
\makeatother
\providecommand*\mcitethebibliography{\thebibliography}
\csname @ifundefined\endcsname{endmcitethebibliography}  {\let\endmcitethebibliography\endthebibliography}{}

\end{document}